\documentclass[letterpaper,journal]{IEEEtran}
\usepackage{graphicx}
\usepackage{subfigure}
\usepackage{float}
\usepackage{amsmath}
\usepackage{epstopdf}
\usepackage{cases}
\usepackage{color}
\usepackage{algorithm}
\usepackage{algpseudocode}
\usepackage{amsmath}
\usepackage{amssymb}
\usepackage{caption}
\usepackage{soul}
\makeatletter
\renewcommand{\maketag@@@}[1]{\hbox{\m@th\normalsize\normalfont#1}}%
\makeatother
\usepackage{stfloats}
\usepackage{cite}
\usepackage{makecell}
\usepackage{multirow}
\usepackage{ulem}
\usepackage{url}

\usepackage{tabularx}  % 自适应列宽
\usepackage{booktabs}  % 三线表样式

\ifCLASSINFOpdf
\else
\fi

\usepackage{caption}
\usepackage{mathtools}
\usepackage{etoolbox}  

\begin{document}
\title{Toward Deeper Environmental Understanding: Event-Level Sensing for Intelligent 6G ISAC}
\author{Haotian Liu,~\IEEEmembership{Graduate Student Member,~IEEE,}
    Zhiqing Wei,~\IEEEmembership{Member,~IEEE,}
    Xingwang Li,~\IEEEmembership{Senior Member,~IEEE,}
    Ruizhong Xu,~\IEEEmembership{Graduate Student Member,~IEEE,} and
    Zhiyong Feng,~\IEEEmembership{Senior Member,~IEEE.}
\thanks{Haotian Liu, Zhiqing Wei(Corresponding Author), Ruizhong Xu, 
and Zhiyong Feng(Corresponding Author) are with Beijing University of Posts and Telecommunications, Beijing, China; Xingwang Li is with Henan Polytechnic University, Jiaozuo, China. 
}
}

\maketitle
\begin{abstract} 
The intelligent evolution of mission-critical networks, such as the Internet of vehicles (IoV) and the low-altitude economy (LAE), requires sixth-generation (6G) networks to move beyond discrete physical parameter estimation toward deeper environmental understanding. 
However, existing integrated sensing and communications (ISAC) studies mainly focus on target-level sensing, which provides fragmented snapshots of the physical world and lacks the behavioral semantic capability to interpret intent. 
This limitation hinders the intelligent evolution of such networks and prevents 6G from acquiring the essential sensing foundation to evolve into an ``intelligent service engine''.
To bridge this gap, ISAC must advance toward event-level sensing, which models continuous-time states to enable persistent recognition and prediction of target intent and behavioral semantics. 
This article presents a comprehensive overview of event-level sensing in 6G ISAC networks. 
We first introduce its fundamental concepts, sensing types, and representative scenarios. 
We then review key enabling techniques across waveform design, target state estimation and tracking, and event recognition. 
Furthermore, focusing on IoV and LAE scenarios, we discuss representative applications of ISAC event-level sensing and the intelligent enhancement of downstream operational functions enabled by event-level information.
Finally, we highlight future research trends and potential directions to further advance ISAC event-level sensing toward intelligent and proactive 6G networks.
\end{abstract}
% \begin{IEEEkeywords}
% Data-driven signal processing,
% integrated sensing and communication (ISAC),
% low-earth-orbit (LEO) constellation,
% cooperative sensing,
% waveform design,
% target tracking.
% \end{IEEEkeywords}

\section{INTRODUCTION}\label{se1}
% 【大背景】通感一体化是6G重点 》 通感一体化的感知还停留在提供目标物理信息的服务的阶段 》 新兴场景需要感知具备主动、提前调控有理解的能力，这要求ISAC感知能力需要从目标级的感知提升到事件级感知 》 ISAC事件级感知有啥优势 》 也有啥挑战（这里的挑战更宏观，全局一些，和本文贡献对应）
% background and motivation
Integrated sensing and communications (ISAC) enables wireless systems with ubiquitous environmental awareness by sharing space, time, frequency, and hardware resources,
and has been recognized as a key enabler for sixth-generation (6G)~\cite{Du2024}. 
However, existing ISAC studies mainly focus on target-level sensing, which characterizes the physical world through instantaneous parameters such as position and velocity~\cite{Luo2025,Liu2026}.
With the intelligent evolution of highly dynamic and complex scenarios, such as autonomous driving and low-altitude economy (LAE), the sensing requirements of the above mission-critical networks have shifted from discrete physical parameter estimation to deeper environmental understanding~\cite{CICT_6G_AI_2024}.

In this context, target-level sensing provides only fragmented snapshots of the physical world, lacking the behavioral semantic capability to interpret intent. 
The lack of behavioral semantics hinders the intelligent evolution of the Internet of vehicles (IoV) and LAE networks, thereby depriving 6G of the essential service foundation to evolve from a passive ``data pipeline'' into an ``intelligent service engine'' as envisioned~\cite{CICT_6G_AI_2024}.
To bridge this gap, ISAC must evolve from target-level to event-level sensing. 
By modeling continuous-time states, event-level sensing enables persistent recognition and prediction of target intent and behavior semantics~\cite{Liu2026}. 
This transition holds significant potential for driving intelligent network transformations across mission-critical networks and accelerates the evolution of 6G intelligent networks~\cite{Liu2026,CICT_6G_AI_2024}.

However, realizing this transition raises three key technical challenges.
\begin{itemize}
    \item \textbf{Waveform design:} Event-level sensing requires waveforms that support robust continuous-time sensing under high mobility while balancing sensing, communication, and RF constraints.
    \item \textbf{Target state estimation and tracking:} Reliable event awareness relies on accurate, robust, and low-complexity continuous target state estimation and tracking in dynamic multi-target environments.
    \item \textbf{Event recognition and inference:} Transforming low-level physical measurements into event semantics requires effective temporal modeling, multi-modal fusion, and efficient inference for complex event evolution.
\end{itemize}

Beyond these technical challenges, the field still lacks a clear conceptual foundation and taxonomy for ISAC event-level sensing, as well as an exploration of its applications and intelligent operation enhancement in IoV and LAE scenarios. These gaps motivate the scope of this article.

Representative target-level ISAC studies have advanced communication- or sensing-centric waveform design~\cite{Fan2026}, physical parameter estimation (e.g., range, velocity, and angle), and target detection, classification, or recognition (e.g., UAV/bird discrimination)~\cite{wang2025,zhang2024,Luo2026}. 
These efforts have significantly improved the sensing capability of ISAC systems at the target level. However, they still mainly concentrate on instantaneous target information and have not yet established an event-level sensing paradigm that captures the continuous evolution of behaviors, intents, and interaction semantics. 
Therefore, a fundamental gap remains between existing target-level sensing studies and the emerging need for event-level sensing in IoV and LAE scenarios.

To the best of our knowledge, this article is the first to introduce ISAC event-level sensing as a new sensing paradigm for 6G networks. 
Specifically, this article presents the concept and taxonomy of ISAC event-level sensing, key enabling techniques, and its applications and intelligent operation enhancement in IoV and LAE scenarios. 
By advancing ISAC from instantaneous parameter awareness to event-level semantic understanding, it aims to support the intelligent transformation of mission-critical networks and promote the evolution of 6G networks into an ``intelligent service engine''. The main contributions are summarized as follows.
\begin{itemize}
    \item \textbf{Foundations of ISAC event-level sensing:} We introduce ISAC event-level sensing as a new sensing paradigm and clarify its fundamental concept, sensing types, representative scenarios, and key events.
    \item \textbf{Key enabling techniques:} This paper analyzes the challenges introduced in waveform design, target state estimation and tracking, and event recognition for event-level sensing. 
    In particular, we present a DFT-s-OFDM-tailored symbol-level joint waveform optimization method for low-peak to average power ratio (PAPR)-constrained ISAC systems and a cross-covariance-enhanced decorrelation parameter estimation approach for robust high-resolution multi-target sensing, and validate their effectiveness through simulations.
    \item \textbf{Applications and intelligent operation enhancement:} We discuss representative applications of ISAC event-level sensing in IoV and LAE scenarios, together with the intelligent enhancement of downstream operational functions enabled by event-level information.
\end{itemize}

\begin{table*}
  \centering
  \caption{Comparison Between Target-Level Sensing and Event-Level Sensing in ISAC}
  \label{tab1}
  \resizebox{0.90\textwidth}{!}{%
  \renewcommand{\arraystretch}{1.5}
  \begin{tabular}{c|l|l}
    \toprule
    \textbf{Aspect} & \textbf{Target-level sensing~\cite{Fan2026,zhang2024,Luo2025,Luo2026}} & \textbf{Event-level sensing~\cite{Liu2026,Fang2024,CICT_6G_AI_2024}} \\
    \midrule
    Sensing objective        & {\begin{tabular}[l]{@{}l@{}}Estimate instantaneous physical parameters  of the \\ target (e.g., distance, velocity, angle)~\cite{Fan2026,Luo2025}\end{tabular}}  & {\begin{tabular}[l]{@{}l@{}}Detect and predict high-level events that may \\ affect network performance or user behavior~\cite{Liu2026}\end{tabular}} \\
    \hline
    Temporal modeling  & {\begin{tabular}[l]{@{}l@{}}Primarily snapshot-based sensing with \\ limited temporal correlation \end{tabular}}   & {\begin{tabular}[l]{@{}l@{}}Explicit temporal modeling of environmental \\ dynamics and event evolution\end{tabular}} \\
    \hline
    Sensing results    & {\begin{tabular}[l]{@{}l@{}}Raw physical measurements or low-level \\ features (e.g., range-Doppler maps, point clouds)  \end{tabular}}    & {\begin{tabular}[l]{@{}l@{}}Semantic event information (e.g., vehicle \\ intent, illegal flight detection) \end{tabular}} \\
    \hline
    Core performance metrics   & {\begin{tabular}[l]{@{}l@{}}Parameter Accuracy (e.g., mean square error,  root mean \\ square error, Cram\'{e}r-Rao lower bound)~\cite{zhang2024}\end{tabular}} & {\begin{tabular}[l]{@{}l@{}}Semantic accuracy and real-time performance \\ (e.g., recognition accuracy, sensing latency, \\ event integrity)~\cite{Fan2026,Liu2026} \end{tabular}} \\
    \hline
    Typical applications      & {\begin{tabular}[l]{@{}l@{}}Localization, tracking, environment mapping, \\ and target detection~\cite{Luo2025,Luo2026} \end{tabular}} & {\begin{tabular}[l]{@{}l@{}}Proactive networking, industrial safety, infrastructure \\ monitoring, and behavior recognition~\cite{Liu2026,Fang2024}\end{tabular}} \\
    \hline
    Implementation complexity   & {\begin{tabular}[l]{@{}l@{}}Relatively mature signal processing pipelines \\ integrated with existing radar or \\ communication frameworks\end{tabular}}  & {\begin{tabular}[l]{@{}l@{}}Requires additional event modeling, temporal \\ inference, and learning-based prediction modules\end{tabular}}  \\
    \hline
    Decision support capability    & {\begin{tabular}[l]{@{}l@{}}Limited support for direct network decision-making \\ due to low-level sensing outputs\end{tabular}}  & {\begin{tabular}[l]{@{}l@{}}Provides actionable event insights that directly \\ support network control and decision processes~\cite{Liu2026,CICT_6G_AI_2024}\end{tabular}} \\
    \hline
    System optimization capability       & {\begin{tabular}[l]{@{}l@{}}Enables reactive optimization after \\ environmental changes are observed \end{tabular}} & {\begin{tabular}[l]{@{}l@{}}Enables proactive system optimization by predicting \\ upcoming environmental events~\cite{Liu2026,CICT_6G_AI_2024}\end{tabular}} \\
    \hline
    Network robustness enhancement   &  {\begin{tabular}[l]{@{}l@{}}Performance may degrade when sudden \\ environmental changes occur\end{tabular}}    & {\begin{tabular}[l]{@{}l@{}}Improves robustness by enabling early awareness \\ and preparation for disruptive events\end{tabular}} \\
    \bottomrule
  \end{tabular}}
\end{table*}

\begin{figure*}[b]
    \centering
    \includegraphics[width=0.90\textwidth]{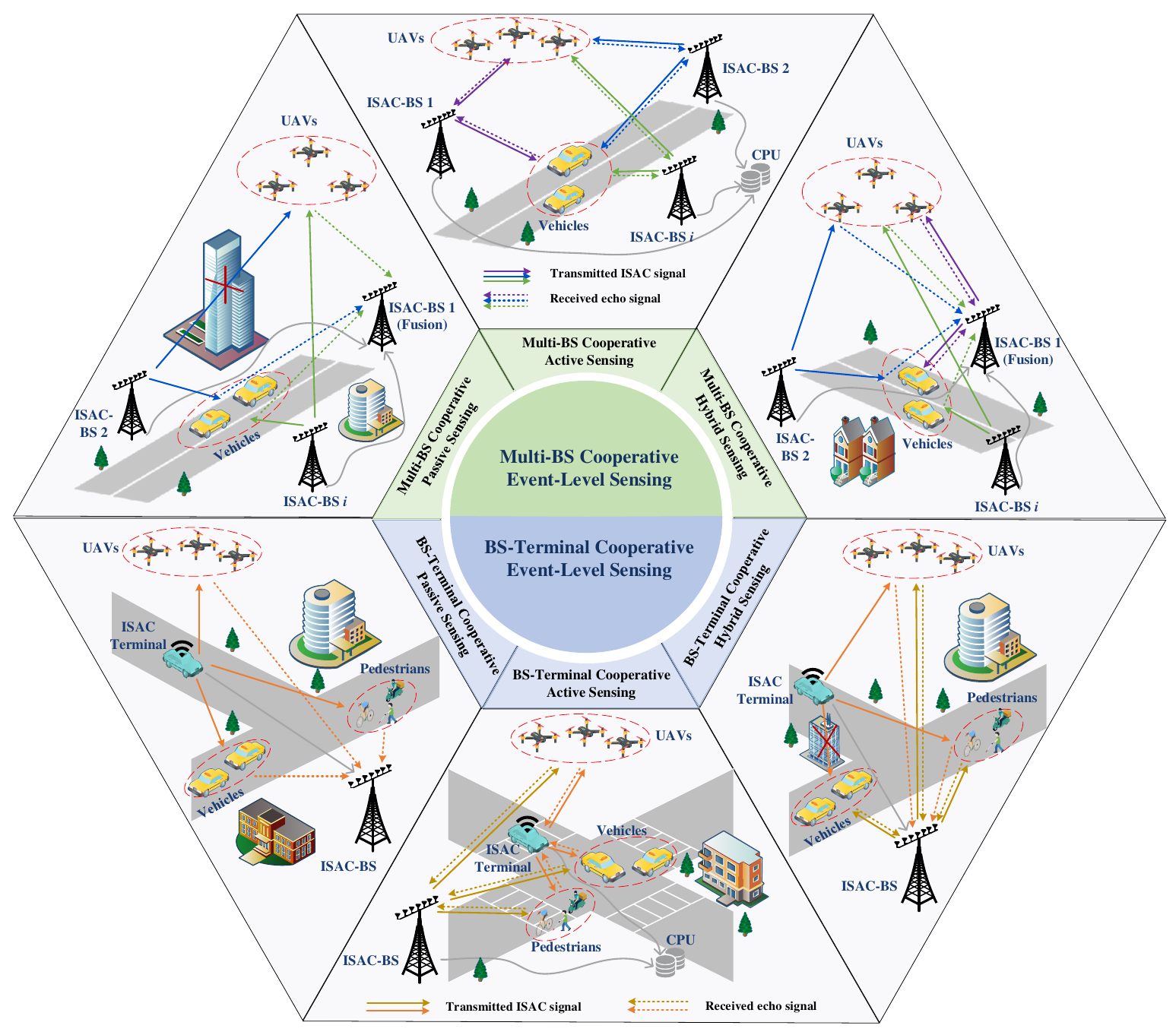}
    \caption{The specific sensing types in cooperative sensing.}
    \label{fig1}
\end{figure*}

\section{FOUNDATION OF ISAC EVENT-LEVEL SENSING}
This section introduces the foundation of ISAC event-level sensing, including the concept, sensing types, representative use cases, and key events.

\subsection{Concept of ISAC event-level sensing}
ISAC event-level sensing refers to leveraging high-rate, multi-dimensional physical state information acquired by ISAC devices, 
combined with temporal modeling and multi-source fusion, 
to recognize or predict individual events or interaction events. 
Unlike traditional target-level sensing, 
which provides fragmented ``snapshots'' of the environment by focusing on instantaneous parameters (e.g., position and velocity), 
event-level sensing emphasizes continuous-time state evolution to capture behavioral semantics and underlying intent. 
Table \ref{tab1} provides a comprehensive comparison between these two paradigms across multiple dimensions.
This evolution provides the critical sensing foundation for endogenous intelligence, 
allowing the network to perceive not just the physical ``where'' of an object, but the semantic ``what'' regarding its future behavior. 
Ultimately, this transformation empowers 6G networks to transcend the role of a passive ``data pipeline'' and evolve into a proactive ``intelligent service engine''~\cite{CICT_6G_AI_2024}.

\subsection{Sensing types of ISAC event-level sensing}
Based on the number of participating nodes, ISAC event-level sensing can be classified into two sensing types~\cite{Wei2024}:
\begin{itemize}
    \item \textbf{Single-node sensing}: In this type, one node independently performs signal transmission, echo reception, signal processing, and event inference. 
    It offers low latency and flexible deployment, making it suitable for localized tasks. 
    However, its performance is often constrained by limited coverage and high susceptibility to blockage. 
    Typical implementations include base station (BS)-based sensing and terminal-based sensing.

    \item \textbf{Cooperative sensing}: In this type, multiple nodes jointly participate in event-level sensing to achieve broader coverage, higher robustness, and more reliable event awareness. 
    As illustrated in Fig.~\ref{fig1}, according to the way sensing information is acquired and fused, cooperative sensing can be further divided into \textit{active}, \textit{passive}, and \textit{hybrid} modes~\cite{Wei2026}:
    \begin{itemize}
        \item \textbf{Active mode}: Each ISAC device independently performs self-transmission and self-reception for event-level sensing.
        \item \textbf{Passive mode}: Multiple ISAC devices transmit sensing signals, while one device receives the reflected or scattered signals for sensing.
        \item \textbf{Hybrid mode}: One device performs self-transmission and self-reception, while also receiving reflected or scattered signals illuminated by other transmitting devices.
    \end{itemize}

    Representative implementation forms of cooperative sensing include:
    \begin{itemize}
        \item \textbf{Multi-BS cooperation}: Multiple BSs jointly participate in event-level sensing. Different BSs can serve as transmitters, receivers, or fusion nodes, thereby exploiting spatial diversity and complementary viewpoints to enhance coverage, blockage resilience, and multi-target resolution.
        
        \item \textbf{BS-terminal cooperation}: BSs and terminals jointly participate in event-level sensing. The BS and terminal can provide complementary uplink and downlink observations, or play different transmitter/receiver roles, enabling multi-scale information acquisition with controllable synchronization and heterogeneous complementarity for real-time sensing.
    \end{itemize}
\end{itemize}

\subsection{Representative use cases and key events}
In complex dynamic environments, event-level sensing serves as the core intelligence that empowers the ISAC network to function as a behavioral semantic awareness hub, as illustrated in Fig.~\ref{fig2}. 
This fundamental shift in sensing capability provides the essential foundation for the intelligent evolution of both IoV and LAE scenarios. 
In IoV scenarios, event-level sensing focuses on individual events, as vehicles act as largely independent traffic participants. 
In LAE scenarios, the strong group dynamics require event-level sensing to extend toward interaction events among multiple UAVs.

\subsubsection{Key events in IoV scenarios}
In dense mixed-traffic IoV scenarios, event-level sensing transforms low-level ISAC observations into individual events. 
This capability improves early awareness of collision risks and traffic conflicts, while supporting safer and more efficient traffic operation.
\begin{itemize}
    \item \textbf{Vehicle intention prediction:} By integrating environmental context, ego-vehicle states, and surrounding vehicle information, ISAC enables the recognition and prediction of intent events such as \textit{overtaking}, \textit{sudden braking}, \textit{sudden acceleration}, and \textit{vehicle stopping}, supporting risk and conflict awareness~\cite{Fang2024}.
    \item \textbf{User behavior recognition:} For vulnerable road users, ISAC can identify high-risk behaviors such as \textit{ghost probe}, \textit{red-light violation}, and \textit{wrong-way riding}, enabling safer cooperative decision-making, especially in non-line-of-sight (NLoS) urban intersections~\cite{Fang2024}.
\end{itemize}

\subsubsection{Key events in LAE scenarios}
In low-altitude logistics and airspace operations, event-level sensing transforms target parameters into behavioral semantic information. 
This capability enables the network to decode coupling relationships among multiple UAVs, ensuring secure and efficient airspace utilization.
\begin{itemize}
    \item \textbf{Violation detection:} Identifying events such as \textit{no-fly zone intrusion}, \textit{abnormal hovering}, and \textit{malicious interference}, supporting dynamic supervision and proactive airspace management~\cite{Wu2025}.
    \item \textbf{Collision-risk and coordination events:} Based on trajectory interaction and behavioral coupling among UAVs, ISAC can detect events such as \textit{trajectory intersection}, \textit{swarm congestion}, and \textit{formation keeping}, enabling early intervention and low-latency coordinated control~\cite{Wu2025}.
\end{itemize}

\begin{figure*}[!b]
    \centering
    \includegraphics[width=0.90\textwidth]{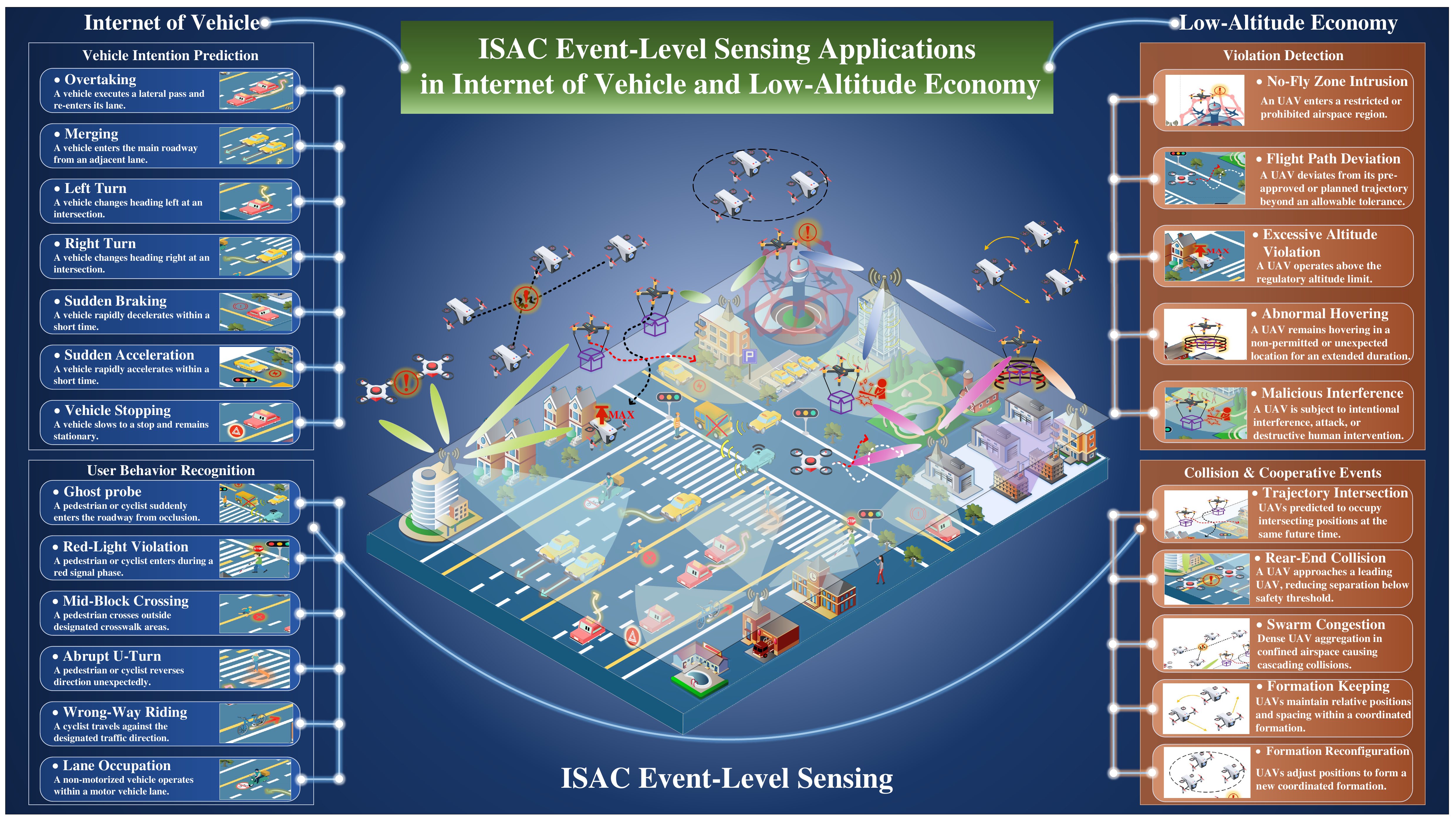}
    \caption{Representative applications and key behavioral events of ISAC event-level sensing.}
    \label{fig2}
\end{figure*}

\begin{figure*}[b]
    \centering
    \includegraphics[width=0.90\textwidth]{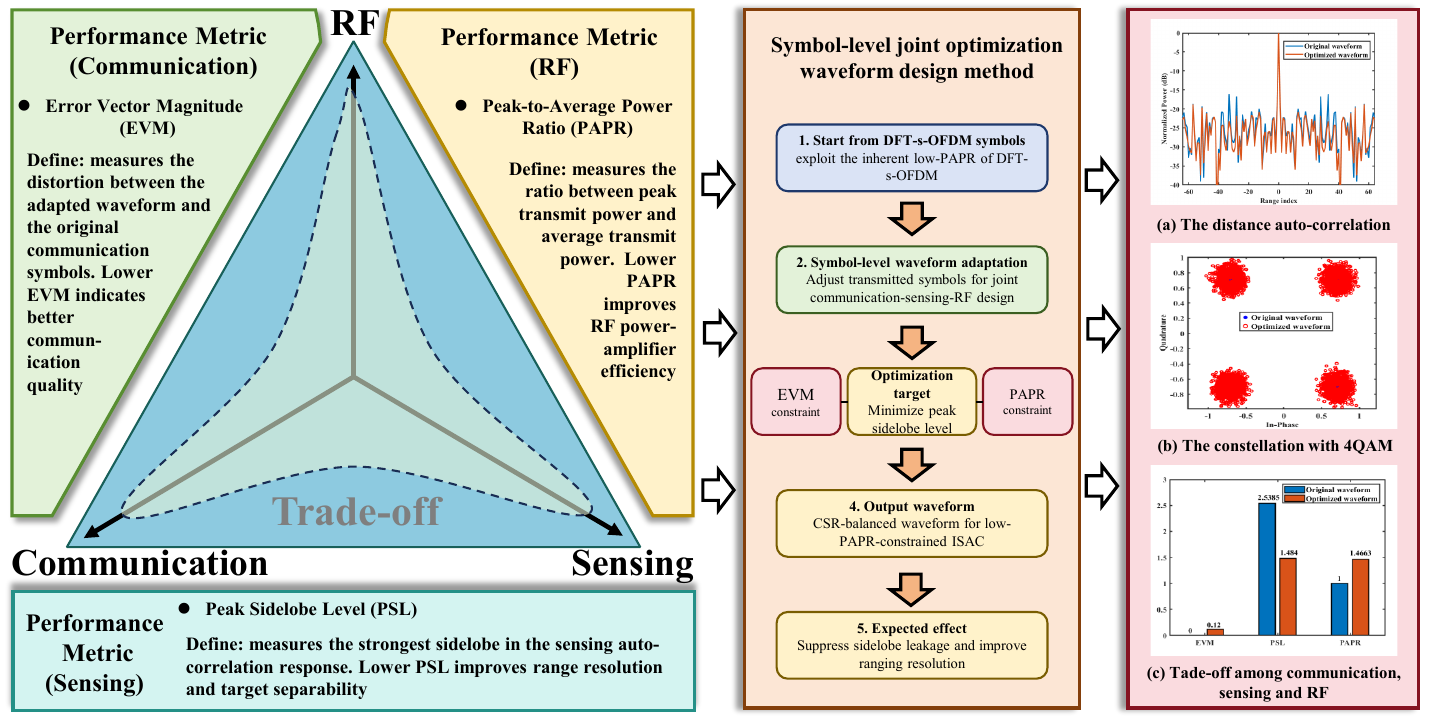}
    \caption{Symbol-level CSR joint optimization waveform design method. Simulation parameters are configured as follows~\cite{Niu2026}:
    the number of transmitted symbols is 256, the EVM constraint coefficient $\varepsilon$ is 0.12, the PAPR constraint value $\gamma$ is 6 dB, and the maximum number of iterations is 15.}
    \label{fig3}
\end{figure*}

\section{KEY ENABLERS FOR ISAC EVENT-LEVEL SENSING}
\label{sec3}
Compared with target-level sensing, event-level sensing introduces new challenges in waveform design, target state estimation and tracking, and event recognition and inference 
due to its strong temporal dynamics and abrupt behavioral transitions. 
To enable the network to evolve into an intelligent service engine, 
these enablers must transform raw physical echoes into high-fidelity event semantic outputs while meeting the stringent requirements for accuracy, integrity, and latency outlined in Table~\ref{tab1}. 
This section outlines key challenges and potential solutions.

\subsection{Waveform design}
Event-level sensing requires enhanced robustness, efficient communication–sensing–RF (CSR) tradeoffs, and adaptive resource control to maintain the semantic accuracy necessary for intent prediction.

\subsubsection{Continuous-time robust waveforms}
Event-level sensing in high-mobility scenarios requires reliable sensing over continuous time, while rapidly time-varying channels may cause severe performance fluctuations.
Waveforms such as orthogonal time frequency space (OTFS) and affine Fourier transform division Multiplexing (AFDM) transform the time-varying channel into a quasi-static two-dimensional representation, thereby maintaining stable sensing features  over time and providing reliable support for continuous trajectory modeling and event evolution analysis.

\subsubsection{CSR waveform design}
Beyond high-resolution sensing and reliable communication, event-level sensing also imposes stringent RF performance requirements due to link stability and hardware constraints. 
Most existing works are built upon conventional OFDM and largely overlook the potential of low-PAPR waveforms. 
Leveraging the inherent low-PAPR property of DFT-s-OFDM, we develop a symbol-level joint optimization framework tailored to DFT-s-OFDM in Fig.~\ref{fig3}, where the transmitted symbols are directly optimized to minimize the PSL under explicit EVM and PAPR constraints. 
Simulation results demonstrate that the proposed method effectively suppresses sidelobe leakage and improves ranging resolution, 
thereby enabling high-resolution sensing for low-PAPR-constrained ISAC systems.

\subsubsection{Multi-resolution and adaptive frame structures}
The requirements on refresh rate and resolution in event-level sensing can change dynamically and abruptly.
To adapt to this characteristic, ISAC systems should support multi-resolution sensing and adaptive frame structures. 
Fortunately, the flexible numerology and dynamic slot scheduling in fifth-generation standards provide essential physical-layer support~\cite{3GPP_38_811_2020}.

\subsection{Target state estimation and tracking}
Event-level sensing shifts the focus from instantaneous state estimation to behavior evolution, requiring higher accuracy, robustness, continuity, and efficiency.
\subsubsection{High-resolution parameter estimation}
\begin{figure*}[!b]
    \centering
    \includegraphics[width=0.90\textwidth]{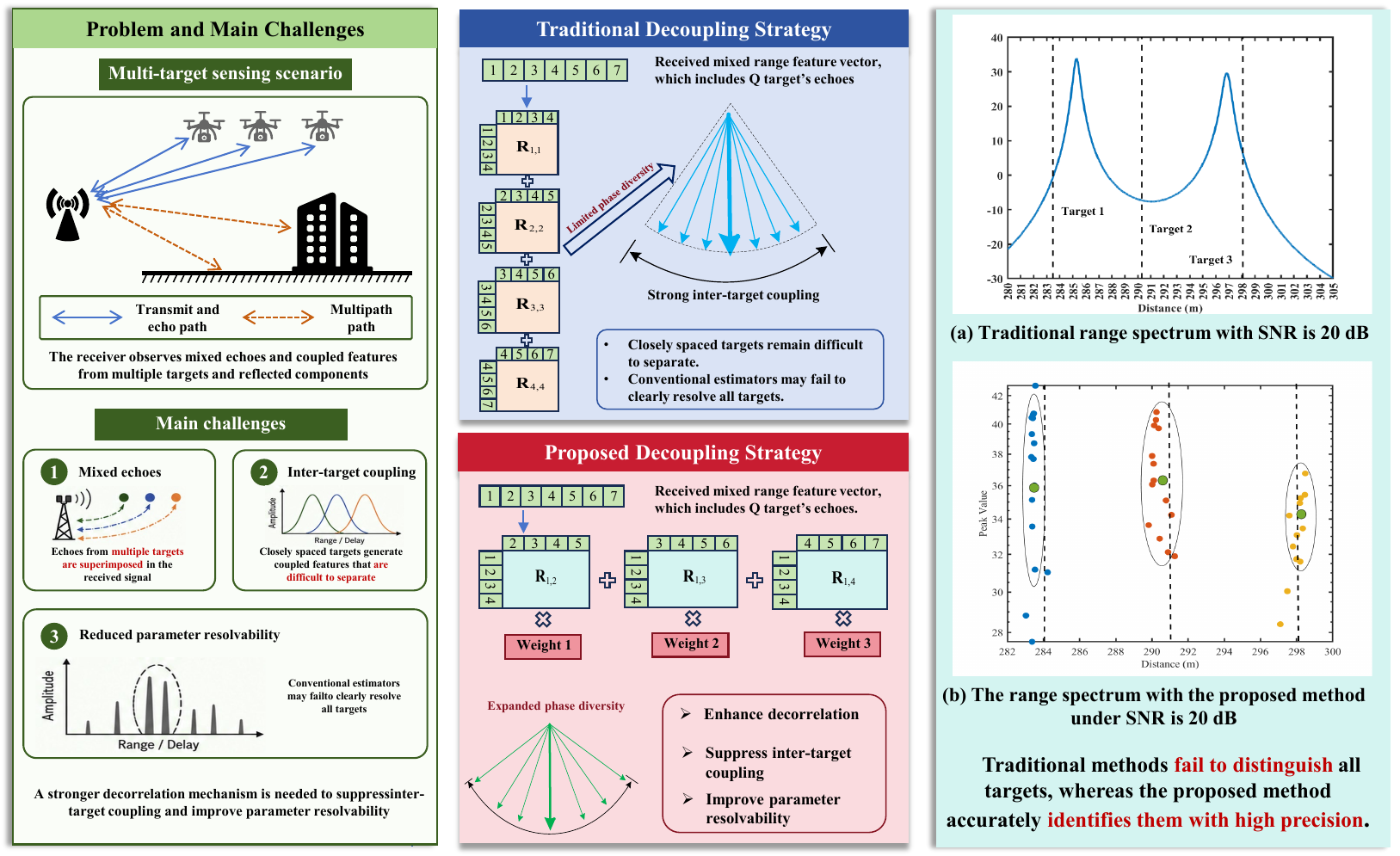}
    \caption{Proposed cross-covariance-enhanced decorrelation parameter estimation method.
    Simulation parameters are configured as follows:
    the carrier frequency is 28 GHz, the bandwidth is 15.36 MHz, the number of subcarriers is 256, the subcarrier spacing is 60 kHz, the length of sub-vector is 128, 
    the ranges of three targets are 284.0633 m, 291.0633 m, and 298.0633 m, respectively.}
    \label{fig4}
\end{figure*}

Event recognition relies on accurate estimation of multi-dimensional target states (e.g., range, velocity, and acceleration).
However, in multi-target scenarios, limited observation diversity and multipath effects often lead to coupled observations and degraded parameter resolvability. 
To improve estimation robustness under such challenging conditions, we propose a cross-covariance-enhanced decorrelation method in Fig.~\ref{fig4}, which can be seamlessly integrated with conventional estimators to suppress inter-target coupling and improve parameter resolvability. 
Simulation results reveal that the proposed approach enables robust and high-resolution multi-target parameter estimation. 
By improving parameter resolvability and suppressing inter-target coupling, this approach provides the high-precision ``raw material'' required to maintain stable and continuous behavior semantics.

\subsubsection{Robust continuous target tracking}
Event-level sensing requires robust and continuous target tracking under nonlinear motion and frequent interactions, 
where conventional constant-velocity or constant-acceleration models become inadequate. 
To this end, random finite set (RFS)-based frameworks provide a systematic way to model target number uncertainty and interactions over time. 
In addition, integrating physics-based motion models with data-driven learning is expected to further improve tracking robustness, continuity, and generalization.

\subsubsection{Low-complexity online processing}
High-dimensional continuous state estimation and multi-target tracking introduce significant computational overhead, limiting real-time deployment. 
One possible approach is to leverage structural priors and low-rank sparse modeling, reducing high-dimensional estimation to low-dimensional optimization, while approximate inference with fast iterative updates enables real-time continuous estimation and tracking for large-scale event-level sensing.

\subsection{Event recognition and inference}
Compared with conventional recognition tasks~\cite{Luo2026}, event recognition faces greater challenges in temporal modeling, multi-modal fusion, and deployment efficiency.

\subsubsection{Temporal feature modeling}
In IoV and LAE scenarios, target states exhibit asynchronous sampling and abrupt changes, 
producing highly non-stationary, multi-timescale event evolution, 
which challenges the stable extraction of discriminative temporal features. 
A promising approach is a hybrid framework combining temporal attention and state-space modeling, capturing both short- and long-term dependencies while maintaining stability and improving the precision and robustness of event evolution characterization.

\subsubsection{Multi-modal information fusion}
Event recognition requires joint reasoning over heterogeneous data, 
including kinematic states (position, velocity, micro-Doppler), 
environmental structures (road topology, airspace layout), and semantic/contextual knowledge (traffic rules, flight restrictions).
To address modality heterogeneity, an integrated framework combining structure-aware modeling, semantic constraints, and attention-based fusion can perform multi-modal reasoning efficiently and interpretably, 
leveraging cross-modal attention and graph-based modeling, 
while incorporating semantic priors to enhance robustness and generalization in complex scenarios~\cite{CICT_6G_AI_2024}.

\subsubsection{Lightweight inference models}
Resource constraints in vehicular and UAV platforms make the direct deployment of complex models impractical.
Event-triggered hierarchical inference provides a promising solution, 
where lightweight edge-side detection is combined with ISAC-device-assisted processing, 
enabling on-demand computation and reduced system overhead while maintaining accuracy.

\section{ISAC EVENT-LEVEL SENSING IN IOV: APPLICATIONS AND INTELLIGENT OPERATION ENHANCEMENT}
\label{se4}
ISAC event-level sensing holds significant potential for driving intelligent network transformation across diverse typical scenarios, particularly in mission-critical IoV and LAE scenarios. 
% This section discusses the applications and intelligent operation enhancement of ISAC event-level sensing in IoV. 
% Specifically, we examine how event-level sensing can be applied in IoV scenarios and how the resulting event semantics can enhance downstream operational functions in a more proactive and intelligent manner. 
% Through these applications, conventional IoV systems can evolve from isolated sensing and reactive operation toward more proactive and intelligent system operation.

\subsection{Integrating event-level sensing into IoV scenarios}
In traditional IoV systems, vehicle intention understanding and vulnerable-road-user behavior recognition mainly rely on onboard or roadside sensors, such as cameras, LiDAR, and mmWave radar~\cite{Fang2024}.
However, these sensors may be limited under NLoS propagation, adverse weather, and long-range sensing conditions.
By leveraging wide-area coverage and networked infrastructure, ISAC can provide large-scale, all-weather, and multi-perspective sensing~\cite{Luo2025}, making it a promising foundation for event-level sensing in IoV.
This capability enables more reliable detection of critical events, such as \textit{ghost probe} and \textit{mid-block crossing}, which may exceed the sensing range of individual vehicles.

Despite its potential, integrating event-level sensing into IoV remains challenging.
A major issue is the temporal inconsistency among heterogeneous data sources: ISAC devices can provide high-rate environmental measurements, while vehicle states, such as velocity, steering angle, and acceleration, are usually reported at lower rates.
Such mismatch may cause state inconsistency and degrade event recognition reliability.
Asynchronous fusion networks can jointly model multi-source temporal data for cross-modal event recognition~\cite{Liu2026}, but their performance still depends on synchronization accuracy and generalization capability.
Future research may further explore event-driven data alignment and semantic-level fusion to improve robustness in complex IoV environments.

\subsection{Intelligent operation enhancement in IoV}
Once applied in IoV scenarios, ISAC event-level sensing provides actionable semantic information that allows ISAC nodes, such as BSs and roadside units (RSUs), to move beyond reactive channel-aware operation. By capturing the underlying intent of vehicles and road users, the IoV infrastructure can enhance downstream operational functions in beam management, resource scheduling, and path planning, as illustrated in Fig.~\ref{fig5}.

\subsubsection{Event-driven beam management}
\begin{figure*}
    \centering
    \includegraphics[width=0.90\textwidth]{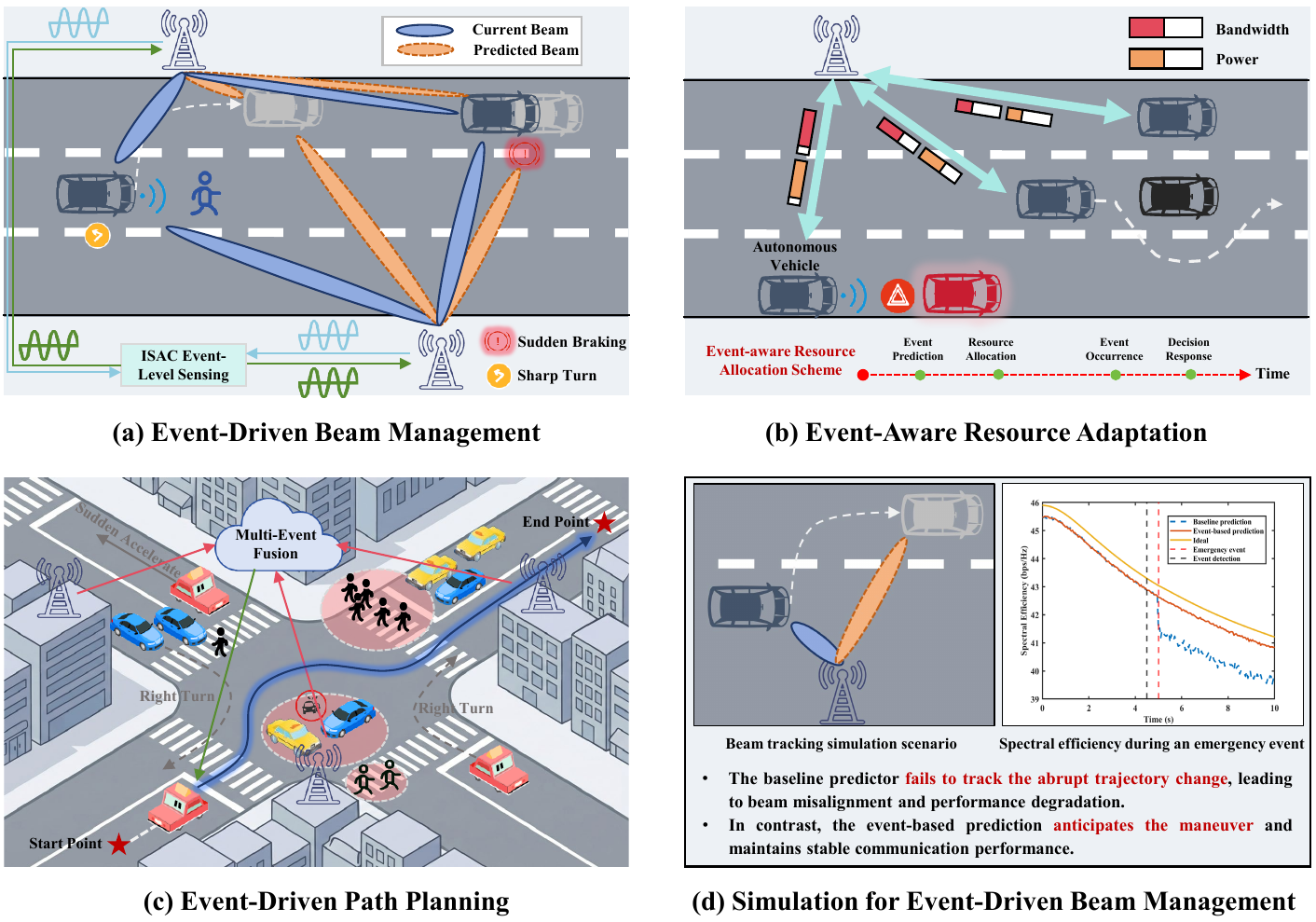}
    \caption{Intelligent operation enhancement enabled by ISAC event-level sensing in IoV}
    \label{fig5}
\end{figure*}

Beam management in IoV must rapidly respond to abrupt motion events such as \textit{sharp turns} and \textit{sudden braking}, which may cause severe link variations.
Conventional methods usually rely on instantaneous channel state information or numerical trajectory extrapolation, and thus may fail to capture abrupt non-linear transitions in vehicle motion.
As shown in Fig.~\ref{fig5}(a), ISAC nodes can leverage event-level sensing to detect behavioral changes and infer motion intent, enabling proactive beam adjustment before the upcoming state transition is reflected in the physical channel.
This allows ISAC-enabled infrastructure to pre-configure candidate beams and maintain more reliable connectivity under abrupt motion changes, as illustrated in Fig.~\ref{fig5}(d).
A key challenge is to balance responsiveness and system complexity, which can be addressed through event-triggered mechanisms and lightweight prediction models.

\subsubsection{Event-aware resource adaptation}
In complex traffic environments, autonomous vehicles must respond to critical events such as \textit{mid-block crossing} and \textit{sudden braking}, imposing latency and reliability requirements on communication links.
As shown in Fig.~\ref{fig5}(b), ISAC event-level sensing enables the identification of high-risk events and supports adjustment of communication resources, including bandwidth, scheduling intervals, and transmit power.
This shifts IoV resource management from long-term statistical allocation to event-driven adaptation, thereby improving the reliability and timeliness of critical data transmission.
However, frequent event triggering in dense scenarios may cause scheduling instability, which can be mitigated by event thresholds and smoothing mechanisms to aggregate and filter events.

\subsubsection{Event-driven path planning}
In dense road networks, limited global traffic awareness often leads to sub-optimal or delayed path planning.
As shown in Fig.~\ref{fig5}(c), event-level sensing enables ISAC infrastructure to capture multi-vehicle behavioral semantics, supporting the evolution of IoV from simple positioning to global coordination for path planning.
Compared with position- and velocity-based methods, event information provides earlier indications of potential risks and improves decision foresight.
Nevertheless, multi-source event data may suffer from temporal, spatial, and reliability inconsistencies, which can affect model stability.
These issues can be alleviated through spatio-temporal alignment and confidence-aware fusion for more reliable risk assessment and decision support.

\section{ISAC EVENT-LEVEL SENSING IN LAE: APPLICATIONS AND INTELLIGENT OPERATION ENHANCEMENT}\label{se5}
Different from IoV scenarios that mainly focus on individual events, LAE is characterized by strong group dynamics and dense interactions among multiple UAVs. 
% This section discusses the applications and intelligent operation enhancement of ISAC event-level sensing in LAE. 
% Specifically, we examine how event-level sensing can be applied in LAE scenarios, especially for interaction events, and how the resulting event semantics can enhance downstream operational functions in low-altitude systems.

\subsection{Integrating event-level sensing into LAE scenarios}
In LAE scenarios, BSs can serve as low-altitude control anchors, while UAVs are usually equipped with lightweight onboard sensors and may not actively share their states. For non-cooperative UAV events shown in Fig.~\ref{fig2}, conventional approaches often depend on cooperative reporting, external surveillance, or multi-stage cloud processing, which may introduce extra latency and system complexity. In contrast, ISAC event-level sensing allows BSs to infer UAV behaviors directly from wireless echoes, enabling low-latency awareness of abnormal flight, trajectory deviation, and potential conflict events.

Nevertheless, integrating event-level sensing into LAE remains challenging. Non-cooperative UAVs may perform rapid three-dimensional maneuvers, causing strong coupling among angle, delay, and Doppler features. Meanwhile, in dense UAV formations or intersecting trajectories, echoes from multiple UAVs may overlap, leading to target ambiguity and unreliable interaction-event recognition. A promising solution is to exploit multi-BS cooperative ISAC for multi-view observation, while incorporating trajectory continuity and interaction consistency into event inference. This can help separate overlapping UAV observations and support more reliable extraction of collision-risk and coordination-event features.

\subsection{Intelligent operation enhancement in LAE}
Once applied in LAE scenarios, ISAC event-level sensing provides BSs with semantic awareness of both individual UAV events and interaction events. Based on these event semantics, the LAE infrastructure can enhance downstream operational functions such as collision avoidance and formation adaptation in a more proactive and intelligent manner.

\subsubsection{Event-aware proactive collision avoidance}
Conventional UAV collision avoidance relies on onboard sensors for local sensing and reactive path adjustment, which is insufficient in dense non-cooperative scenarios due to limited awareness of interaction risks.
In contrast, ISAC event-level sensing enables BSs to detect multi-UAV behaviors and predict conflicts before they occur, transforming the LAE network into a coordinated safety overseer.
With event awareness, UAVs can trigger avoidance strategies proactively rather than relying on local observations.
However, the uncertainty and abruptness of maneuver events may affect avoidance stability.
This can be mitigated by event confidence evaluation and short-term interaction modeling, enabling smoother avoidance execution while maintaining safety and efficiency.

\subsubsection{Event-driven efficient formation adaptation}
In low-altitude logistics and cooperative missions, UAVs often operate in formations, where conventional formation control mainly relies on state exchange among UAVs or through BSs but lacks semantic understanding of cooperative behaviors.
As a result, events such as member deviation or load imbalance may be detected late, leading to delayed adaptation.
With ISAC event-level sensing, BSs can identify UAV behaviors and interactions in real time, enabling the LAE network to autonomously monitor formation health.
This supports event-triggered reconfiguration and dynamic task redistribution for improved mission efficiency.
To avoid instability caused by frequent adjustments, event-triggered hierarchical control can be introduced to balance responsiveness and global stability, promoting a more self-adaptive LAE system.

\section{Future Trends}
This section discusses the future research trends and potential solutions for ISAC event-level sensing.

% \subsection{City-scale event sensing for environmental awareness}
% Future ISAC-empowered wireless networks may evolve into large-scale sensing infrastructures capable of monitoring dynamic events across complex urban environments. 
% By leveraging dense deployments of BSs and their continuous sensing capabilities, ISAC event-level sensing can enable city-scale awareness of traffic flow, crowd dynamics, infrastructure anomalies, and low-altitude aerial activities. 
% This large-scale event sensing provides a unified sensing view of urban dynamics, enabling coordinated understanding of spatially distributed events and more anticipatory urban management. 

% However, achieving reliable city-scale event sensing requires addressing several challenges, 
% including high computing power and low latency processing center, real-time event detection from massive sensing streams, 
% and efficient fusion of heterogeneous sensing information across distributed network nodes.

\subsection{Space-air-ground integrated event-level sensing networks}
Beyond city-scale sensing, ISAC event-level sensing may further evolve toward space-air-ground integrated sensing networks.
By integrating terrestrial BSs, aerial sensing platforms, and low-earth-orbit satellites, event-level sensing can be extended from urban areas to regional or even global coverage~\cite{Wei2026}.
This capability can support large-scale disaster monitoring, environmental change detection, and wide-area airspace and maritime surveillance.

Moreover, space-based event-level sensing may enable the monitoring of deep-space and orbital events, such as satellite anomalies and space debris dynamics.
However, the heterogeneity of sensing platforms introduces new challenges in synchronization, sensing-resolution alignment, and multi-layer data fusion across different network segments.

\subsection{Event-aware AI-native wireless networks}
As wireless networks evolve toward AI-native architectures with endogenous intelligence, event-level sensing provides a critical foundation for environment-aware network autonomy.
By capturing high-level environmental dynamics, such as mobility transitions, blockage events, and infrastructure changes, wireless networks can anticipate disruptive events and adapt resources, topologies, and service policies in advance.
This shifts network operation from reactive optimization to intent-driven proactive management.
Moreover, continuous event observation allows networks to refine operational policies through online learning, supporting self-evolving wireless systems in complex scenarios such as dense urban mobility and large-scale aerial networks.

However, realizing this vision requires new learning frameworks, such as network operation large models (NOLMs)~\cite{CICT_6G_AI_2024}, capable of integrating real-time sensing data, communication measurements, and long-term environmental knowledge while maintaining scalability and real-time responsiveness.

% \subsection{Event-level sensing-assisted covert transmission}

% Future ISAC event-level sensing may open new opportunities for covert transmission in security-sensitive wireless networks. 
% Conventional covert communication usually assumes the existence of a warden, also known as Willie, and focuses on reducing the detectability of transmission behavior by exploiting channel uncertainty, artificial noise, or conservative resource control~\cite{Bash2013,Chen2023}. 
% However, in highly dynamic 6G networks, potential wardens may be unknown, mobile, and behaviorally adaptive. 
% Therefore, the network first needs to infer whether suspicious nodes exist, how they move, and when they may create high transmission-exposure risks.

% Event-level sensing provides a semantic basis for this problem. 
% By continuously tracking targets and modeling their behavior patterns, ISAC networks can identify abnormal mobility, suspicious proximity, persistent following, or unusual interaction patterns that may indicate potential wardens. 
% Such event-level awareness can further guide covert transmission decisions, including when to transmit, which beam direction to use, how much power to allocate, and which time-frequency resources are less exposed. 
% However, realizing this vision requires joint designs of event recognition, suspicious-node inference, and covert resource allocation under sensing uncertainty and intelligent wardens.

\subsection{Event-driven digital twins of physical environments}
ISAC event-level sensing opens new opportunities for intelligent digital twins with enhanced situational awareness and reasoning capabilities.
By capturing high-level environmental events rather than only raw physical measurements, ISAC-enabled networks can provide richer semantic representations of the physical world.
This allows digital twins to better understand system dynamics, predict emerging risks, and support closed-loop decision-making.
As highlighted in the 6G vision~\cite{CICT_6G_AI_2024}, event-driven twins can facilitate internal closed-loop verification and external closed-loop feedback, enabling applications such as abnormal drone detection, long-term urban planning, and integrated communication-sensing-control systems in smart factories~\cite{Wei2026iot}.

Enabling such semantic digital twins requires further research on event abstraction, multi-source event fusion, and high-fidelity synchronization between the physical environment and its virtual counterpart.

\section{Conclusion}
In this article, ISAC event-level sensing is presented as a promising paradigm shift toward the intelligent evolution of mission-critical networks. 
We provide a comprehensive overview of this emerging paradigm from the perspectives of conceptual foundation, key enabling techniques, applications and intelligent operation enhancement, and future research trends. 
Specifically, we clarified the concept and sensing types of ISAC event-level sensing, and discussed its representative use cases in IoV and LAE scenarios. 
We then reviewed the main technical challenges and enabling techniques across waveform design, target state estimation and tracking, and event recognition. 
Moreover, we discussed representative applications of ISAC event-level sensing in IoV and LAE scenarios, together with the intelligent enhancement of downstream operational functions enabled by event-level information.
Finally, we outlined several future directions, such as event-aware AI-native wireless networks, event-level sensing-assisted covert transmission, and event-driven digital twins.
We hope this article can stimulate further interest in ISAC event-level sensing and contribute to the evolution of 6G networks toward an ``intelligent service engine''.

% In this article, ISAC event-level sensing is proposed as a pivotal paradigm shift to drive the intelligent evolution of mission-critical vertical networks. 
% This article provides a comprehensive overview of the fundamental concepts, enabling technologies, and empowerment pathways for this new sensing paradigm.
% We first defined the concept and sensing types of event-level sensing and explored its representative use cases in IoV and LAE scenarios. 
% Subsequently, we investigated the technical challenges and key enabling techniques across waveform design, target state estimation and tracking, and event recognition. 
% Then, we discussed the implementation of event-level sensing and its ways to drive the intelligent transformation of mission-critical vertical networks.
% In addition, we highlighted several future research trends, such as event-aware AI-native wireless networks and event-driven digital twins. 
% This article may inspire further research interest in ISAC event-level sensing, 
% which holds significant potential to support the evolution of 6G networks into an ``intelligent service engine''.

% reference
\bibliographystyle{IEEEtran}
\bibliography{reference}

\end{document}